\documentclass[aps,prb,reprint,superscriptaddress,floatfix]{revtex4-2}

\usepackage{ulem}
\usepackage{graphicx}
\usepackage{dcolumn}
\usepackage{bm}
\usepackage{color}
\usepackage{comment}
\usepackage{amsmath}
\usepackage{amssymb}
\usepackage{hyperref}

\begin{document}

\title{Compensation Domain Walls in Ferrimagnets as Spin-Torque Oscillator Candidates}

\author{Matteo~Fettizio}
\email{mfettizio@icmab.es}
\affiliation{Institut de Ciència de Materials de Barcelona (ICMAB-CSIC), Carrer dels Til.lers, 08193 Cerdanyola del Vallès, Spain}

\author{Can~Onur~Avci}
\email{cavci@icmab.es}
\affiliation{Institut de Ciència de Materials de Barcelona (ICMAB-CSIC), Carrer dels Til.lers, 08193 Cerdanyola del Vallès, Spain}

\date{\today}

\begin{abstract}
Spin-torque nano-oscillators are promising nanoscale microwave sources, yet their performance is often limited by narrow operational ranges and sensitivity to device parameters. Here, we numerically investigate the dynamics of compensation domain walls in ferrimagnetic systems driven by spin-transfer torque as a microwave source candidate. These walls arise in structures with adjacent regions of different compensation character, where the competition between exchange and Zeeman energies under an applied magnetic field stabilizes a distinct magnetic texture. We demonstrate steady current-driven rotation of compensation domain walls over a broad range of current densities within the model. To elucidate the underlying mechanism, we develop a collective-coordinate model and derive an analytical expression for the oscillation frequency, which quantitatively reproduces the dependence on current and magnetic field observed in micromagnetic simulations. Our results identify compensation domain walls as a candidate platform for tunable spin-torque nano-oscillators with intrinsic field-induced confinement.
\end{abstract}

\maketitle

\section{Introduction}

Nanoscale auto-oscillators are promising components for communication, sensing, and neuromorphic technologies. Magnetic nanodevices are attractive due to their compactness, nonvolatility, and coupling between charge and spin degrees of freedom \cite{Brataas2012,Locatelli2014}.

Spintronic oscillators based on spin-transfer-torque (STT) and spin-orbit-torque (SOT) have been widely explored as nanoscale microwave sources \cite{Kiselev2003,Rippard2004,Chen2016,zahedinejad2020two,houshang2016spin, haidar2019single}. In STT oscillators, a spin-polarized current compensates magnetic damping and sustains steady-state precession \cite{Slonczewski1996,Berger1996}, while in SOT devices the torque originates from current-induced spin accumulation in adjacent heavy metals \cite{Liu2012,Manchon2019}. These mechanisms enable GHz-frequency emission in magnetic nanopillars and nanocontacts \cite{Kiselev2003,Rippard2004}, but can suffer from high threshold currents, linewidth broadening, and restricted operational ranges \cite{Chen2016,Slavin2009,Mazraati2018}.

An alternative approach exploits domain walls (DWs) confined in thin films with perpendicular magnetic anisotropy. Pinned DWs can sustain steady oscillatory dynamics under spin torque \cite{bisig2009tunable,voto2017pinned}. Their operating range, however, requires sufficiently strong confinement to prevent depinning while allowing sustained internal rotation, making their dynamics sensitive to geometry and disorder \cite{bisig2009tunable,voto2017pinned}.

Ferrimagnets provide a qualitatively different platform due to the antiferromagnetic coupling between inequivalent sublattices, which can lead to magnetization or angular-momentum compensation \cite{Kim2017,Caretta2018}. In structures containing adjacent regions with different compensation character, exchange--Zeeman competition under a sufficiently strong magnetic field stabilizes a compensation DW separating distinct equilibrium configurations \cite{blanco2015tuning,Deb2021tunable,Wang2021probing, shiino2026giantnonlinearhalleffect,shiino2026tunable}. Unlike conventional ferromagnetic DWs governed by exchange--anisotropy balance, these walls exhibit distinct static and dynamical properties.

Here, we investigate STT-driven dynamics of compensation DWs. We develop a collective-coordinate model, derive analytical expressions for the oscillation frequency and its upper operating limit, and compare them with micromagnetic simulations. We further show that the threshold current can be intrinsically reduced, identifying compensation DWs as a candidate platform for tunable spin-torque nano-oscillators.

\section{System's description}

We consider a structure composed of two adjacent ferrimagnetic layers with different magnetization compensation temperatures, $T_{\mathrm{CM},1}$ and $T_{\mathrm{CM},2}$. The operating temperature $T$ is chosen such that $T_{\mathrm{CM},1}<T<T_{\mathrm{CM},2}$, so that the two layers have opposite compensation character. Equivalently, their net magnetizations have opposite orientations when referred to the same magnetic sublattice. Magnetic anisotropy is neglected in order to isolate the exchange--Zeeman mechanism responsible for the formation and confinement of the compensation DW. 

Although this configuration can in principle be realized in any ferrimagnet exhibiting a compensation point, we focus here on rare-earth--transition-metal (RE--TM) ferrimagnets, which represent one of the most common classes of compensated ferrimagnets \cite{Sala2022Ferrimagnetic}. This choice also simplifies the treatment of STT: owing to the localized nature of the RE $4f$ electrons, their exchange coupling to the conduction electrons is weak, allowing us to assume that the current-induced spin torque acts predominantly on the TM sublattice. 

In the strong-exchange limit, the antiferromagnetically coupled RE and TM sublattices remain rigidly collinear and can be described by a single unit vector $\mathbf{m}=(\sin\theta\cos\phi,\sin\theta\sin\phi,\cos\theta)$ representing the TM spin direction. Each position is characterized by a local effective magnetization $M_{\mathrm{S}}=M_{\mathrm{TM}}-M_{\mathrm{RE}}$, whose sign determines the local compensation character. For the numerical simulations, we neglect in-plane spatial variations and discretize the system along $z$ as a one-dimensional chain of exchange-coupled spins. Taking the interface between the two regions at $z=0$ and considering a RE-dominated bottom layer, we write

\begin{equation}
M_{\mathrm{S}}(z)=
\begin{cases}
-M_2, & -L/2<z<0,\\
M_1, & 0<z<L/2.
\end{cases}
\end{equation}

Each TM spin is ferromagnetically coupled to the adjacent ones. To account for the opposite compensation character of the two regions, the effective field acting on the $i$-th spin is written as

\begin{equation}
\mathbf{H}_{\mathrm{eff},i}
=
\mathrm{sgn}(M_{\mathrm{S},i})\,\mathbf{H}_{\mathrm{ext}}
+
\frac{J}{\mu_0|M_{\mathrm{S},i}|}
\left(
\mathbf{m}_{i-1}+\mathbf{m}_{i+1}
\right),
\end{equation}

where $J>0$ is the nearest-neighbor ferromagnetic exchange-energy density of the discrete model, related to the continuum exchange stiffness ($A$) by $J=2A/a_{\mathrm{Fe}}^2$, with $a_{\mathrm{Fe}}$ the Fe--Fe spacing. The factor $\mathrm{sgn}(M_{\mathrm{S},i})$ accounts for the reversal of the net magnetization relative to the TM sublattice across the interface, whereas the exchange contribution includes the term $|M_{\mathrm{S},i}|$ as it only depends on the direction of the TM spins. 

\section{Field-induced compensation domain wall}

We first investigate the static properties of the system in the absence of STT, with simulation parameters listed in Supplemental Material \cite{SupplementalMaterial}.

We initialize the magnetic state at zero field with all TM spins pointing along $-z$. This uniform configuration corresponds to the exchange-dominated ground state, in which the spatial exchange interaction favors the alignment of all TM spins. For the case considered here, $M_2>M_1$, the corresponding net magnetization points along the positive $z$ direction. Since magnetic anisotropy is neglected, there is no preferred orientation of the uniform magnetic state, and all configurations obtained by a global rotation of the aligned spins are degenerate at zero field.

\begin{figure}[!b]
\includegraphics[width=\columnwidth]{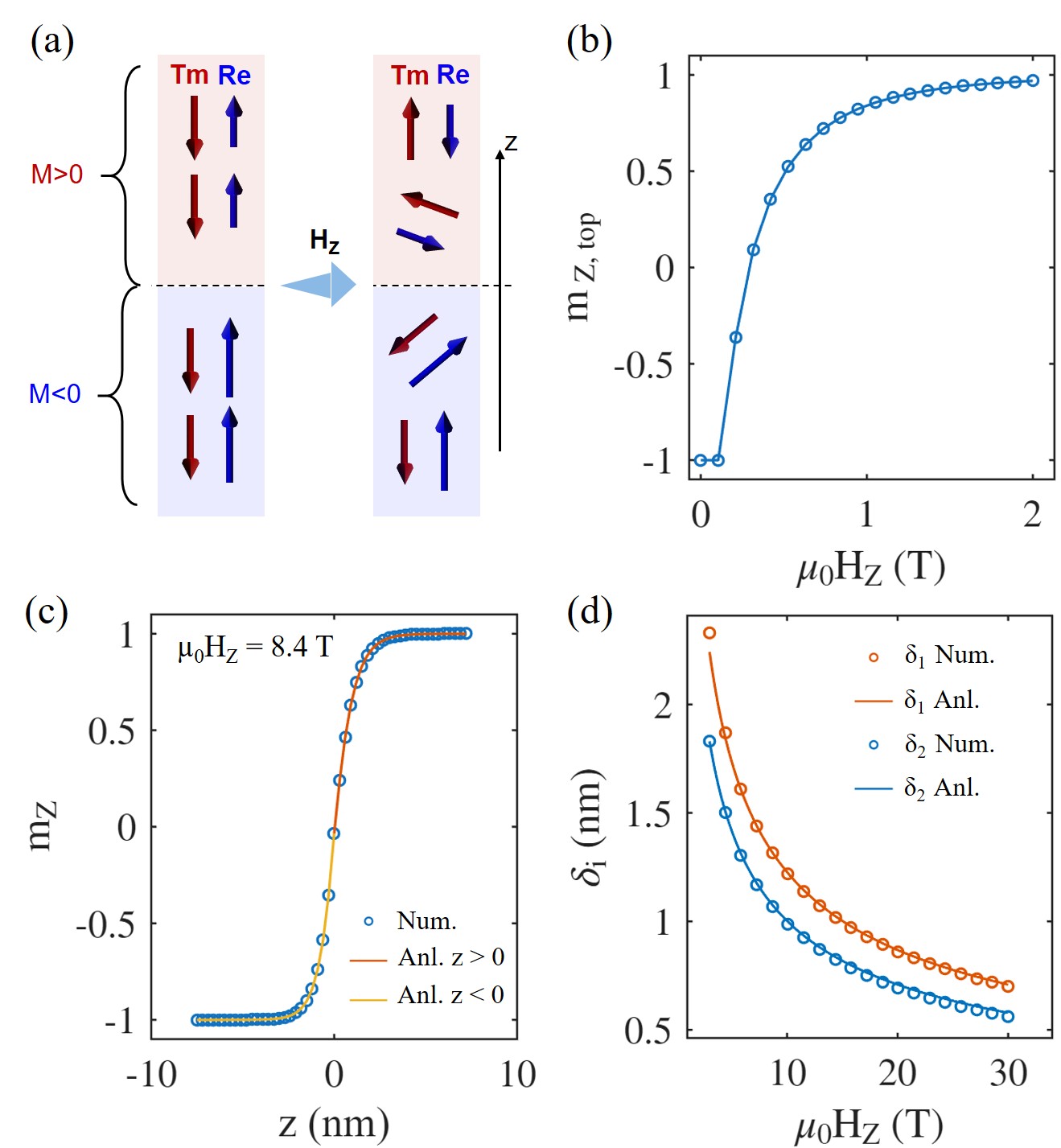}
\caption{(a) Schematic illustration of the formation of a compensation DW under an applied field $H_{\rm Z}$ in a ferrimagnet composed of two regions with different compensation character. (b) $m_{\rm Z}$ of the top spin as a function of $H_{\rm Z}$. (c) Simulated spatial profile of $m_{\rm Z}$ (blue dots), together with fits obtained from Eq.~\eqref{eq_profile} (orange and yellow curves). (d) Field dependence of $\delta_1$ (orange dots) and $\delta_2$ (blue dots), together with the analytical prediction of Eq.~\eqref{eq:dw_delta} (solid lines).
}
\label{Fig_CDW}
\end{figure}

Starting from this initial configuration, $H_{\rm Z}>0$ is progressively applied. As the field is increased, the competition between exchange and Zeeman energies eventually destabilizes the uniform state. The spins in the top region progressively rotate to minimize their Zeeman energy, leading to the formation of a compensation DW separating regions with different equilibrium orientations, as illustrated in Fig.~\ref{Fig_CDW}(a). The transition is further evidenced in Fig.~\ref{Fig_CDW}(b), which shows the computed $z$ component of $\mathbf{m}$, $m_{\rm Z}$, of the top spin as a function of $H_{\rm Z}$ for a chain of 50 spins. The top spin provides a convenient probe of the magnetic configuration, since its complete reversal indicates that the DW is fully developed. Because no magnetic anisotropy is included, the in-plane magnetization direction is degenerate, and we therefore focus solely on $m_{\rm Z}$. This approximation simplifies the analytical treatment without changing the underlying exchange--Zeeman mechanism; a finite uniaxial perpendicular anisotropy would mainly increase the transition field and reduce the DW width. The DW starts forming at $\mu_0H_{\rm Z}=75$~mT and gradually narrows with increasing field.

For fields above which $m_{\rm Z, top}\simeq 1$, corresponding to a fully formed compensation DW, we analyze the DW profile. The numerically obtained $m_{\rm Z}(z)$ profile for $\mu_0H_{\rm Z} = 8.4$ T is shown in Fig.~\ref{Fig_CDW}(c) (blue dots). In contrast to conventional ferromagnetic DWs, where the profile results from the balance between exchange and anisotropy in a homogeneous system, here the wall is governed by exchange--Zeeman competition in a spatially inhomogeneous medium. As a result, a modified analytical description is required.

Minimization of the total energy leads to a piecewise-defined solution for the DW profile. For instance, the polar angle reads (details of the derivation can be found in Supplemental Material \cite{SupplementalMaterial}):

\begin{equation}
\resizebox{\columnwidth}{!}{$\displaystyle
\theta(z)=
\begin{cases}
\pi - 4\,\arctan\!\left[
\tan\!\left(\dfrac{\pi-\theta_0}{4}\right)\,e^{z/\delta_2}
\right], & -L/2<z<0, \\[6pt]
4\,\arctan\!\left[
\tan\!\left(\dfrac{\theta_0}{4}\right)\,e^{-z/\delta_1}
\right], & 0<z<L/2,
\end{cases}
$}
\label{eq_profile}
\end{equation}

where $\theta_0$ is the polar angle at the boundary between the two regions, corresponding to $\theta_0 = 2\,\arctan\!\left(\sqrt{M_2/M_1}\right)$. $\delta_1$ and $\delta_2$ are the characteristic widths in the two regions which, in contrast to the conventional case,  depend on the applied field:
\begin{equation}
\delta_{\rm i} = \sqrt{\frac{2A}{\mu_0 M_{\rm i} H_{\rm Z} }}.
\label{eq:dw_delta}
\end{equation}

The analytical solution provides an excellent fit to the numerical data when applied separately to the two sides of the wall, as shown in Fig.~\ref{Fig_CDW}(c) (orange and yellow curves). This confirms the validity of the piecewise description. As the applied field increases, the Zeeman energy increasingly favors alignment within each region, leading to a progressive reduction of the DW width. This behavior is quantified in Fig.~\ref{Fig_CDW}(d), where the extracted values of $\delta_1$ (orange dots) and $\delta_2$ (blue dots) are shown together with the analytical prediction (solid curves), demonstrating good agreement and confirming the field dependence described above.

\section{Collective coordinate model}

To describe the DW dynamics, we employ a collective coordinate approach, where the magnetic texture is reduced to a small set of dynamical variables. This framework has been widely used to model DW motion under applied fields and spin torques, providing accurate descriptions of both steady-state and dynamical regimes \cite{Schryer1974,Thiaville2005,Tatara2008}.

The magnetization dynamics is governed by the Landau--Lifshitz--Gilbert equation (LLG) including STT in the Zhang--Li form 
\cite{Zhang2004},

\begin{equation}
\frac{\partial \mathbf{m}}{\partial t} =
-\gamma\mu_0 \, \mathbf{m} \times \mathbf{H}_{\mathrm{eff}}
+ \alpha \, \mathbf{m} \times \frac{\partial \mathbf{m}}{\partial t}
- u_{\mathrm{STT}}(z)\, \frac{\partial \mathbf{m}}{\partial z},
\end{equation}

where $\gamma$ is the gyromagnetic ratio, $\alpha$ is the Gilbert damping and $u_{\mathrm{STT}}(z)$ is the spin-transfer velocity. For the analytical treatment, we neglect non-adiabatic STT.

The spin-transfer velocity $u_{\mathrm{STT}}(z)$ depends explicitly on position through the local magnetization, $u_{\mathrm{STT}}(z) = (P j_0\mu_B)/(eM_1)$ for $0<z<L/2$ and $u_{\mathrm{STT}}(z) = (P j_0 \mu_B)/(eM_2)$  for $-L/2<z < 0$, where $P$ is the spin polarization, $j_0$ is the current density, $\mu_B$ is the Bohr magneton, $e$ is the elementary charge, and $M_1$ and $M_2$ denote the magnitude of the effective magnetization in the two regions. $u_{\mathrm{STT}}$ depends on $|M_{\rm S}|$ because STT acts mainly on the TM sublattice but its strength depends on the net magnetic moment \cite{Sala2022Ferrimagnetic}. 

In conventional ferromagnetic systems, the dynamics is described in terms of the DW center and the azimuthal angle $\phi$. In the present case, however, the DW center is pinned at $z=0$ due to the competition between exchange and Zeeman interactions; therefore, deformations of the DW have to be described in terms of variations of $\theta_0$ and $\delta_{1,2}$, which, together with $\phi$, become the collective coordinates of the system.

The equations of motion for the generic collective coordinate $\xi_{\rm i}$ take the following form \cite{tretiakov2008dynamics}:
\begin{equation}
G_{\rm ij}\,\dot{\xi}_{\rm j} - \gamma\partial_{\xi_{\rm i}}U - F_{\xi_{\rm i}}^{\rm STT} - \Gamma_{\rm ij}\,\dot{\xi}_{\rm j} = 0,
\label{eq_ccm}
\end{equation}
where $G_{\rm ij}$ and $\Gamma_{\rm ij}$ are the elements of the gyrotropic and dissipation tensors, respectively; $\partial_{\xi_{\rm i}}U$ is the $\xi_{\rm i}$ derivative of the free magnetic energy per cross-sectional area and $F_{\xi_{\rm i}}^{\rm STT}$ is the STT force coefficient.

We focus on a steady-state regime in which $\phi$ is the only dynamical variable. This requires the presence of an effective pinning force that suppresses fluctuations of the remaining degrees of freedom. While in conventional pinned DW oscillators confinement is typically engineered
(e.g., via notches), here pinning arises intrinsically: displacement of the DW
away from the interface increases the Zeeman energy in one of the two regions,
thereby confining the DW near $z=0$. This behavior is captured by the forces
$-\partial_{\xi_{\rm i}}U$ (with $\xi_{\rm i}\neq\phi$), which, upon
linearization around equilibrium with the other coordinates held fixed, yield
terms proportional to $(\xi_{i,\mathrm{eq}}-\xi_{\rm i})$, where
$\xi_{i,\mathrm{eq}}$ is the equilibrium value of $\xi_{\rm i}$, corresponding
to a spring-like restoring force (Supplemental Material \cite{SupplementalMaterial}). It is therefore possible to consider a regime in which $\theta_0$, $\delta_1$, and $\delta_2$ remain stationary in time, i.e., $\dot{\theta}_0=0$ and $\dot{\delta}_{1,2}=0$. Imposing these conditions in Eq.~\eqref{eq_ccm} yields 

\begin{equation}
\dot{\phi} = -\frac{F_{\phi}^{\mathrm{STT}}}{\Gamma_{\phi\phi}}.
\label{eq_phidot_general}
\end{equation}

Assuming a DW centered at $z=0$, the coefficients are obtained by integration over $z \in [-L/2,L/2]$. Defining $s_0=\sin(\theta_0/2)$, $c_0=\cos(\theta_0/2)$, $c_+ = \cos[\theta(L/2)/2]$, $s_- = \sin[\theta(-L/2)/2]$ and $c_- = \cos[\theta(-L/2)/2]$, we obtain (details of derivation can be found in Supplemental Material \cite{SupplementalMaterial})
\begin{equation}
\Gamma_{\phi\phi}
= \frac{4\alpha}{3}\left[
M_2\delta_2\left(s_-^3 - s_0^3\right)
+ M_1\delta_1\left(c_+^3 - c_0^3\right)
\right],
\end{equation}
\begin{equation}
F_{\phi}^{\mathrm{STT}} 
= \frac{2Pj_0\mu_B}{e}
\left(
c_-^2 - c_+^2
\right).
\end{equation}

To obtain an experimentally relevant expression for $\dot{\phi}$, we consider the small STT limit, in which $\theta_0$ remains close to its equilibrium value $\theta_{0,\mathrm{eq}} = 2\,\arctan\!\big(\sqrt{M_2/M_1}\big)$, and $\delta_1$ and $\delta_2$ remain close to the equilibrium values given by Eq.~\eqref{eq:dw_delta}. In addition, we assume a fully developed DW, so that we can set $\theta(L/2) = 0$ and $\theta(-L/2) = \pi$.

Under these conditions, and taking $\theta(L/2)\simeq 0$ and $\theta(-L/2)\simeq \pi$, one obtains an explicit expression for $\dot{\phi}$ in terms of the system parameters:
\begin{widetext}
\begin{equation}
\dot{\phi}
=
\frac{3Pj_0 \mu_{\rm B}}{2e\alpha \sqrt{\dfrac{2A}{\mu_0 H_{\rm Z}}}
\left[
\sqrt{M_1}\!\left(1-\dfrac{M_1^{3/2}}{(M_1+M_2)^{3/2}}\right)
+
\sqrt{M_2}\!\left(1-\dfrac{M_2^{3/2}}{(M_1+M_2)^{3/2}}\right)
\right]}.
\label{eq_phidot}
\end{equation}
\end{widetext}

The expression is valid for $H_{\rm Z},M_1,M_2>0$ and the sense of rotation is determined by the sign of $j_0$. The magnitude of this result resembles the expression for conventional DW oscillators, $|\dot{\phi}| = |u_{\mathrm{STT}}|/(\alpha\delta)$ \cite{bisig2009tunable,voto2017pinned}, where $\delta$ is the DW parameter, while explicitly incorporating the spatial inhomogeneity of the system.

In typical PMA systems, oscillations arise from the rotation of the DW between Bloch and Néel configurations, which requires overcoming an energy barrier associated with in-plane anisotropy and Dzyaloshinskii–Moriya interaction if present. As a result, a finite threshold current is needed to sustain the dynamics. While in such systems geometrical engineering is generally required to reduce the energy barrier and thus the threshold current density, in the present system the threshold can be intrinsically minimized. By employing cylindrical geometries and materials with negligible in-plane anisotropy, such as amorphous RE--TM ferrimagnets, the energy barrier for internal rotation can be strongly suppressed, potentially enabling the onset of oscillations at low current densities.

Regarding the maximum operating current, it is determined by the largest STT strength for which a steady-state solution of Eq.~\eqref{eq_ccm} exists under the condition $\dot{\xi}_{i\neq\phi}=0$. Such a solution exists as long as the steady-state values of the coordinates $\xi_{\rm i}$ satisfy

\begin{equation}
-\,G_{i\phi}\frac{F_{\phi}^{\mathrm{STT}}}{\Gamma_{\phi\phi}}
= \gamma\,\partial_{\xi_{\rm i}}U.
\label{eq:steady_state_collective_balance}
\end{equation}

In typical pinned DW oscillators, Eq. \eqref{eq:steady_state_collective_balance} is satisfied as long as  the current-induced drive can be balanced by the pinning force. Beyond this limit, the stationary solution
ceases to exist and the DW moves away from the notch.
As discussed in Section \ref{sec_V}, steady-state oscillations persist up to current densities sufficiently large to significantly distort the DW, such that the zero-STT profile of Eq.~\eqref{eq_profile} is no longer valid. Consequently, the steady-state solutions for $\theta_0$, $\delta_1$, and $\delta_2$ cannot be used to reliably estimate the maximum current density. An alternative approach is therefore introduced in Section \ref{sec_V}.

\section{Oscillations under low STT}

\begin{figure}[b]
\includegraphics[width=\columnwidth]{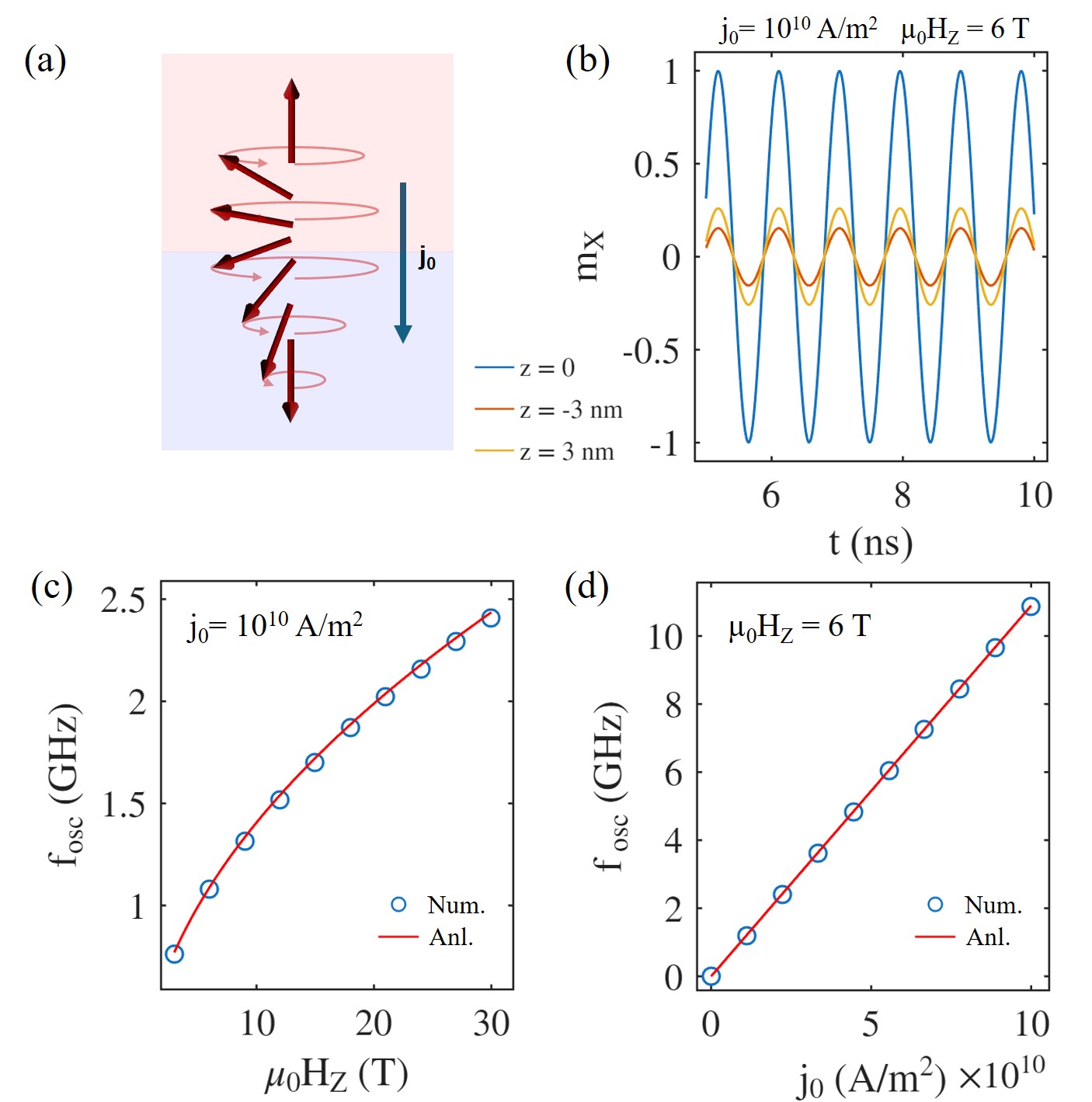}
\caption{
(a) Schematic illustration of the DW dynamics under STT upon application of the current density $j_0$. (b) Time evolution of $m_{\rm X}$ for the spins located at $z=0$ (blue curve), $z=3\,\mathrm{nm}$ (yellow curve), and $z=-3\,\mathrm{nm}$ (orange curve). (c) $f_{\rm osc}$ as a function of $\mu_0 H_{\rm Z}$: micromagnetic simulations (blue dots) and analytical prediction from Eq.~\eqref{eq_phidot} (red curve) for $j_0 = 10^{10}\,\mathrm{A/m^2}$. (d) $f_{\rm osc}$ as a function of $j_0$: micromagnetic simulations (blue dots) and analytical prediction from Eq.~\eqref{eq_phidot} (red curve) at $\mu_0 H_{\rm Z}=6\,\mathrm{T}$.
}
\label{Fig_moderateSTT}
\end{figure}

To validate the developed model, we simulate the response of the system introduced in the previous sections under a constant STT, as a function of the external field $H_{\rm Z}$ and the current density $j_0$. Since the in-plane components of $\mathbf{m}$ are now also of interest, a small field component of order mT along the $x$ direction is introduced to lift their degeneracy. As illustrated in Fig.~\ref{Fig_moderateSTT}a, under the action of STT the DW is expected to undergo a rigid rotation around the $z$ axis, along which $j_0$ flows.
As a consequence, the in-plane components of the DW magnetization oscillate with an amplitude that depends on the local value of $m_{\rm Z}$, which remains constant in time.

This behavior is confirmed in Fig.~\ref{Fig_moderateSTT}(b), where the time dependence of the $x$ component of the magnetization, $m_{\rm X}(t)$, is shown for different spins at $\mu_0H_{\rm Z} = 6$ T and $j_0 = 10^{10}$ A/m$^2$. As expected, the oscillation amplitude is maximal for spins located near the interface between the two regions (blue curve, $z=0$) and progressively decreases away from it (orange and yellow curves). The different oscillation amplitudes observed at equal distances from $z=0$ are a consequence of the asymmetric DW profile (Fig.~\ref{Fig_CDW}(c)).

The oscillation frequency $f_{\rm osc} = |\dot{\phi}|/(2\pi)$ as a function of $\mu_0 H_{\rm Z}$ is shown in Fig.~\ref{Fig_moderateSTT}(c) (blue symbols), together with the analytical prediction of Eq.~\eqref{eq_phidot} (red line) for $j_0 = 10^{10}\,\mathrm{A/m}^2$. Figure~\ref{Fig_moderateSTT}(d) displays $f_{\rm osc}$ versus $j_0$ in the range $0 \le j_0 \le 10^{11}\,\mathrm{A/m}^2$ at fixed $\mu_0 H_{\rm Z} = 6\,\mathrm{T}$, again in comparison with Eq.~\eqref{eq_phidot} (red line). The excellent agreement between simulations and the collective-coordinate model substantiates the validity of the approach in this regime and supports the assumption of low STT.

\section{Oscillations under strong STT}
\label{sec_V}

\begin{figure}[t]
\includegraphics[width=\columnwidth]{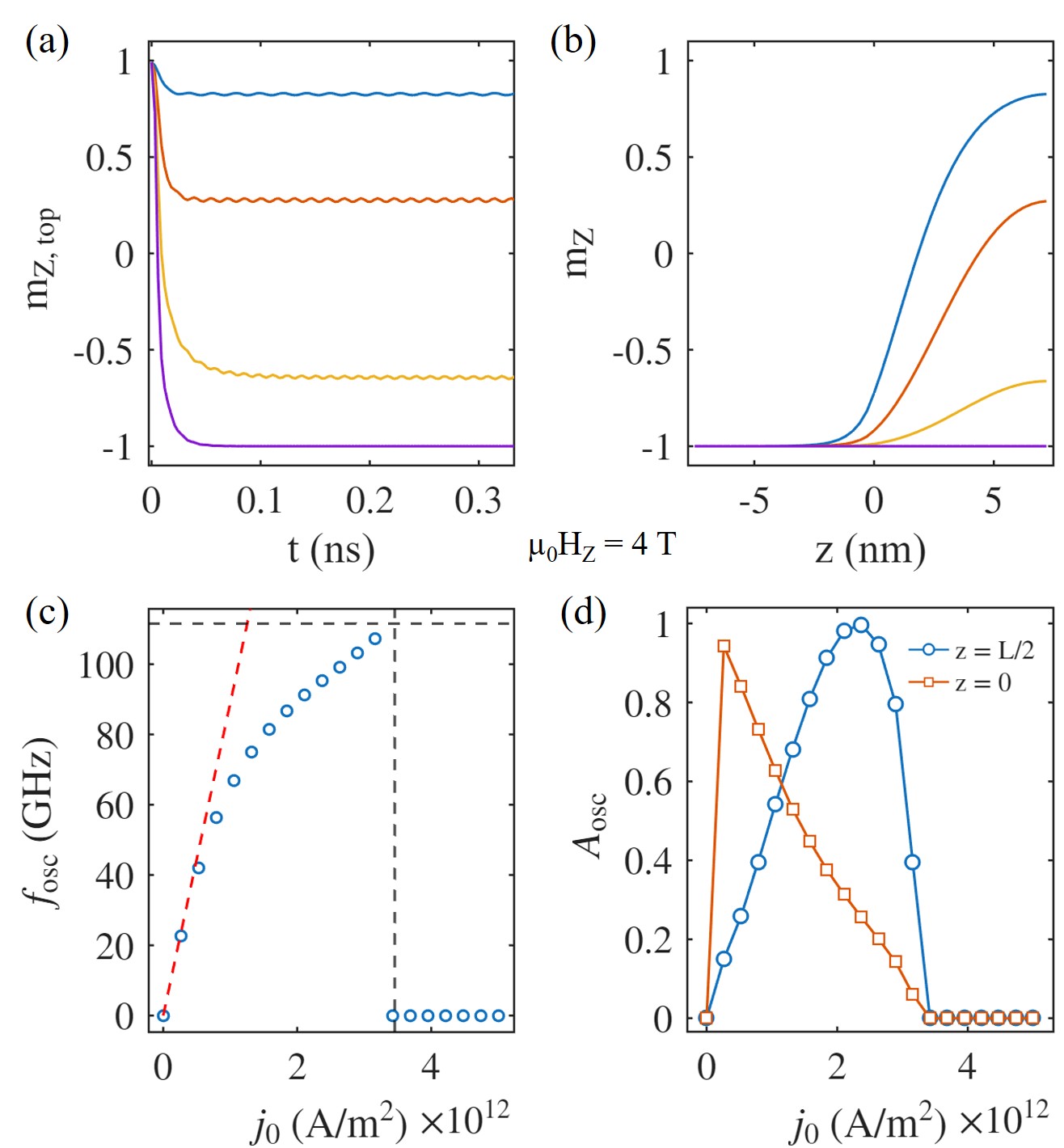}
\caption{
(a) Time evolution of $m_{\rm Z,top}$ after application of STT for different values of $j_0$ at $\mu_0H_{\rm Z}=4\,\mathrm{T}$. (b) Corresponding DW spatial profiles. (c) Oscillation frequency $f_{\rm osc}$ as a function of $j_0$: micromagnetic simulations (blue dots) and low-STT prediction from Eq.~\eqref{eq_phidot} (red dotted curve). The predicted $j_{\rm max}$ and $f_{\rm max}$ are indicated by the vertical and horizontal black dotted lines, respectively. (d) $A_{\rm osc}$ as a function of $j_0$ for the top spin (blue curve) and the central spin (orange curve).
}
\label{Fig_strongSTT}
\end{figure}

To investigate the upper boundary for steady-state oscillations and the effect of strong STT, we consider current densities of order $10^{12}\,\mathrm{A/m}^2$. We first let the system relax in the absence of STT and then apply a constant current (the timing parameters are given in Supplemental Material \cite{SupplementalMaterial}). Figure~\ref{Fig_strongSTT}(a) shows the time evolution of $m_{\rm Z,top}$ at $\mu_0H_{\rm Z}=4\,\mathrm{T}$ for increasing $j_0$. Immediately after the application of STT, a rapid transient drives $m_{\rm Z,top}$ toward $-1$. The magnitude of this variation increases with $j_0$ until a fully collinear configuration is stabilized and the oscillations are suppressed. The residual oscillations at lower current densities arise from the small field canting along $x$, introduced to break the in-plane magnetization degeneracy.

Figure~\ref{Fig_strongSTT}(b) shows that increasing STT progressively distorts the DW away from its equilibrium profile of Eq.~\eqref{eq_profile}. Because the STT is nonconservative, this distorted steady state is governed by the LLG equation rather than by energy minimization. The deformation is mainly reflected in the increase of the top-boundary angle $\theta_+$ toward $\pi$, allowing us to identify $j_{\rm max}$ with the condition $\theta_+\to\pi$, at which the DW collapses.

To derive an approximate balance equation for $\theta_+$ analogous to Eq.~\eqref{eq:steady_state_collective_balance}, an explicit ansatz for the distorted DW profile is required. For simplicity, for $z>0$ we assume a linear profile of the form
$\theta(z)=\pi-(2z/L)(\pi-\theta_+)$, with $\theta=\pi$ for $z<0$.

Within this linear-profile approximation, and neglecting exchange in the high-field force balance (Supplemental Material \cite{SupplementalMaterial}), the limit $\pi-\theta_+\equiv\Delta_+\to0$ yields

\begin{equation}
j_{\rm max}
=
\frac{e}{P\mu_B}
\left(
\frac{\alpha\gamma\mu_0H_{\rm Z}M_1L}{3}
\right).
\end{equation}

The corresponding limiting oscillation frequency is

\begin{equation}
f_{\rm max} = (\gamma\mu_0H_{\rm Z})/2\pi,
\end{equation}

which corresponds to the Larmor frequency. The expression for $j_{\rm max}$ is proportional to the Zeeman energy per unit cross-sectional area of the top region through the factor $\mu_0H_{\rm Z}M_1L$, showing that stronger Zeeman confinement requires a larger STT to collapse the DW.

Figure~\ref{Fig_strongSTT}(c) shows the oscillation frequency of the central spin as a function of $j_0$ up to $5\times10^{12}\,\mathrm{A/m^2}$ at $\mu_0H_{\rm Z}=4\,\mathrm{T}$. The low-STT prediction of Eq.~\eqref{eq_phidot} reproduces the simulations at moderate current densities but breaks down for $j_0\gtrsim10^{12}\,\mathrm{A/m^2}$, where the progressive distortion of the DW invalidates the equilibrium-profile approximation. In contrast, the predicted $j_{\rm max}$ closely matches the onset of oscillation quenching, while the corresponding $f_{\rm max}$ is also in good agreement with the simulated limiting frequency.

The estimate of $j_{\rm max}$ is quantitatively reliable only as long as the strongly distorted DW can be approximated by the assumed linear profile. Further simulations show that outside a certain range of $H_{\rm Z}$ and $L$ this approximation becomes less accurate. Nevertheless, it provides a reliable order-of-magnitude estimate of the upper current boundary and captures its dependence on the relevant system parameters.

Finally, since the oscillation amplitude $A_{\rm osc}$ is directly relevant for the output power of the oscillator, we examine its dependence on $j_0$ at fixed $\mu_0H_{\rm Z}=4\,\mathrm{T}$. As shown in Fig.~\ref{Fig_strongSTT}(d), the central-spin amplitude is largest at low current densities, whereas the top-spin amplitude initially increases with $j_0$, reaches a maximum at $j_0=2.37\times10^{12}\,\mathrm{A/m^2}$, corresponding to the largest canting of the top spin, and then decreases as $j_{\rm max}$ is approached.

\section{Experimental implementation}

A direct realization could employ compositionally graded metallic RE--TM ferrimagnets, such as GdCo, GdFe, or GdFeCo, where the compensation point can be tuned through composition \cite{Suzuki2023,damerio2024magnetoresistive}. A nanopillar could therefore be designed with upper and lower regions on opposite sides of compensation, reproducing the spatially varying effective magnetization of our model. A current flowing perpendicular to the film would provide the STT required to trigger the oscillation, while the resulting dynamics could be detected electrically through anisotropic magnetoresistance.

A complementary implementation could use ferrimagnetic insulators such as REIGs. Magnon currents generated by spin-Hall injection or Joule heating in an adjacent heavy metal could transfer angular momentum to the compensation DW \cite{Yan2011}, with detection through spin Hall magnetoresistance \cite{Nakayama2013} or exchange coupling to a metallic ferromagnet. The main practical limitation in both cases is the external field required to stabilize the DW, which could potentially be replaced by a built-in exchange or magnetostatic bias.

\section{Conclusions}

In conclusion, we showed that exchange--Zeeman competition in ferrimagnets with spatially varying compensation can stabilize intrinsically confined DWs capable of steady-state rotation under STT. A collective-coordinate model captures the low-STT dynamics and predicts the oscillation frequency, while an extended treatment of the strongly distorted DW provides estimates of $j_{\rm max}$ and $f_{\rm max}$, in good agreement with simulations.

Experimentally, the proposed mechanism could be realized in compositionally graded metallic RE--TM ferrimagnets driven by electronic STT, or in ferrimagnetic insulators driven by magnonic spin currents, with electrical detection through magnetoresistive effects. These results identify compensation DWs as a candidate platform for tunable spin-torque nano-oscillators with intrinsic confinement and potentially low-current operation. The main practical limitation is the magnetic field required to stabilize the DW. Field-free operation could potentially be achieved through built-in exchange or magnetostatic bias, while efficient electrical readout and synchronization between coupled oscillators represent natural directions for future development.

\begin{acknowledgments}
We acknowledge funding from the European Research Council (ERC) under the European Union's Horizon 2020 research and innovation programme through the MAGNEPIC project (Grant Agreement No.~949052). We acknowledge financial support from the State Investigation Agency through the Severo Ochoa Programme for Centres of Excellence in R\&D (CEX2023-001263-S), and funding from the MUST project (PCI2023-143400/MCIU/AEI/10.13039/501100011033), co-funded by the European Union. We thank Takayuki Shiino for insightful discussions.
\end{acknowledgments}

\section*{Data Availability}
The data supporting the findings of this study are available from the corresponding author upon reasonable request.

\bibliography{Bibliography}



\end{document}


\setcounter{equation}{0}
\renewcommand{\theequation}{S.\arabic{equation}}
\title{Supplemental Material for ``Compensation Domain Walls in Ferrimagnets as Spin-Torque Oscillator Candidates''}

\author{Matteo~Fettizio}
\affiliation{Institut de Ciència de Materials de Barcelona (ICMAB-CSIC), Carrer dels Til.lers, 08193 Cerdanyola del Vallès, Spain
}

\author{Can~Onur~Avci}
\email{cavci@icmab.es}
\affiliation{Institut de Ciència de Materials de Barcelona (ICMAB-CSIC), Carrer dels Til.lers, 08193 Cerdanyola del Vallès, Spain
}

\maketitle
\tableofcontents

\section{Equilibrium profile from energy minimization}

We consider a one-dimensional magnetization texture
$\mathbf m(z)=
\left(
\sin\theta(z)\cos\phi,
\sin\theta(z)\sin\phi,
\cos\theta(z)
\right)$, 
with uniform azimuthal angle $\phi$. The magnetic energy per cross-sectional area is
\begin{equation}
U=
\int dz
\left[
A(\theta')^2
-\mu_0 M_{\rm S}(z)H_{\rm Z}\cos\theta
\right],
\end{equation}
where $\theta'=d\theta/dz$ and
\begin{equation}
M_{\rm S}(z)=
\begin{cases}
-M_2, & z<0,\\
M_1, & z>0.
\end{cases}
\end{equation}

Minimization of the energy functional yields
\begin{equation}
2A\theta''-\mu_0M_1H_{\rm Z}\sin\theta=0,
\qquad z>0,
\end{equation}
and
\begin{equation}
2A\theta''+\mu_0M_2H_{\rm Z}\sin\theta=0,
\qquad z<0.
\end{equation}

Integrating once gives
\begin{equation}
A(\theta')^2
=
\mu_0M_1H_{\rm Z}(1-\cos\theta),
\qquad z>0,
\end{equation}
and
\begin{equation}
A(\theta')^2
=
\mu_0M_2H_{\rm Z}(1+\cos\theta),
\qquad z<0.
\end{equation}

Introducing
\begin{equation}
\delta_i=
\sqrt{
\frac{2A}{\mu_0H_{\rm Z}M_i}
},
\label{eq:si_widths}
\end{equation}
the solutions satisfying
\begin{equation}
\theta(-\infty)=\pi,
\qquad
\theta(+\infty)=0,
\qquad
\theta(0)=\theta_0,
\end{equation}
are
\begin{equation}
\theta(z)=
\begin{cases}
\displaystyle
\pi
-
4\arctan\!\left[
\tan\!\left(\frac{\pi-\theta_0}{4}\right)
e^{z/\delta_2}
\right],
& z<0,
\\[4mm]
\displaystyle
4\arctan\!\left[
\tan\!\left(\frac{\theta_0}{4}\right)
e^{-z/\delta_1}
\right],
& z>0.
\end{cases}
\end{equation}

Finally, minimizing the total energy with respect to $\theta_0$ gives
\begin{equation}
M_1\delta_1\sin\left(\frac{\theta_{0,\rm eq}}{2}\right)
=
M_2\delta_2\cos\left(\frac{\theta_{0,\rm eq}}{2}\right),
\end{equation}
which yields
\begin{equation}
\tan\left(\frac{\theta_{0,\rm eq}}{2}\right)
=
\sqrt{\frac{M_2}{M_1}}.
\end{equation}

\section{Derivation of Thiele coefficients}

Considering finite integration limits $z\in[-L/2,L/2]$, we define the endpoint angles $\theta_-=\theta(-L/2), \qquad \theta_+=\theta(L/2)$
and use the shorthand notation
$s_0=\sin\frac{\theta_0}{2},\quad
c_0=\cos\frac{\theta_0}{2},\quad
s_\pm=\sin\frac{\theta_\pm}{2},\quad
c_\pm=\cos\frac{\theta_\pm}{2}.$
The dissipative tensor associated with the azimuthal coordinate is
\begin{equation}
\Gamma_{\phi\phi}
=
\alpha\int_{-L/2}^{L/2}
|M_{\rm S}(z)|
\left|
\frac{\partial\mathbf m}{\partial\phi}
\right|^2 dz .
\end{equation}
Since
\begin{equation}
\frac{\partial\mathbf m}{\partial\phi}
=
\sin\theta\,\mathbf e_\phi,
\end{equation}
we have
\begin{equation}
\left|
\frac{\partial\mathbf m}{\partial\phi}
\right|^2
=
\sin^2\theta .
\end{equation}
Therefore
\begin{equation}
\Gamma_{\phi\phi}
=
\alpha M_2\int_{-L/2}^{0}\sin^2\theta\,dz
+
\alpha M_1\int_{0}^{L/2}\sin^2\theta\,dz .
\end{equation}

For $z>0$ we use
\begin{equation}
\theta'=-\frac{2}{\delta_1}\sin\frac{\theta}{2},
\qquad
dz=-\frac{\delta_1}{2}
\frac{d\theta}{\sin(\theta/2)}.
\end{equation}
Thus
\begin{align}
\int_{0}^{L/2}\sin^2\theta\,dz
&=
\int_{\theta_0}^{\theta_+}
4\sin^2\frac{\theta}{2}
\cos^2\frac{\theta}{2}
\left[
-\frac{\delta_1}{2}
\frac{d\theta}{\sin(\theta/2)}
\right] \nonumber\\
&=
2\delta_1
\int_{\theta_+}^{\theta_0}
\sin\frac{\theta}{2}
\cos^2\frac{\theta}{2}
\,d\theta .
\end{align}
Using $u=\cos(\theta/2)$, $du=-(1/2)\sin(\theta/2)d\theta$, we obtain
\begin{equation}
\int_{0}^{L/2}\sin^2\theta\,dz
=
\frac{4\delta_1}{3}
\left(c_+^3-c_0^3\right).
\end{equation}

For $z<0$,
\begin{equation}
\theta'=-\frac{2}{\delta_2}\cos\frac{\theta}{2},
\qquad
dz=-\frac{\delta_2}{2}
\frac{d\theta}{\cos(\theta/2)}.
\end{equation}
Hence
\begin{align}
\int_{-L/2}^{0}\sin^2\theta\,dz
&=
\int_{\theta_-}^{\theta_0}
4\sin^2\frac{\theta}{2}
\cos^2\frac{\theta}{2}
\left[
-\frac{\delta_2}{2}
\frac{d\theta}{\cos(\theta/2)}
\right] \nonumber\\
&=
2\delta_2
\int_{\theta_0}^{\theta_-}
\sin^2\frac{\theta}{2}
\cos\frac{\theta}{2}
\,d\theta .
\end{align}
Using $v=\sin(\theta/2)$, $dv=(1/2)\cos(\theta/2)d\theta$, we get
\begin{equation}
\int_{-L/2}^{0}\sin^2\theta\,dz
=
\frac{4\delta_2}{3}
\left(s_-^3-s_0^3\right).
\end{equation}

Therefore,
\begin{equation}
\Gamma_{\phi\phi}
=
\frac{4\alpha}{3}
\left[
M_2\delta_2\left(s_-^3-s_0^3\right)
+
M_1\delta_1\left(c_+^3-c_0^3\right)
\right].
\end{equation}

The projection of the adiabatic spin-transfer torque onto the azimuthal coordinate is
\begin{equation}
F_{\phi}^{\rm STT}
=
\frac{P\mu_B j_0}{e}
\int_{-L/2}^{L/2}
\frac{\partial\mathbf m}{\partial\phi}
\cdot
\left(
\mathbf m\times\frac{\partial\mathbf m}{\partial z}
\right)dz .
\end{equation}
Since
\begin{equation}
\frac{\partial\mathbf m}{\partial z}
=
\theta'\mathbf e_\theta,
\qquad
\mathbf m\times\frac{\partial\mathbf m}{\partial z}
=
\theta'\mathbf e_\phi,
\end{equation}
and
\begin{equation}
\frac{\partial\mathbf m}{\partial\phi}
=
\sin\theta\,\mathbf e_\phi,
\end{equation}
the integrand becomes
\begin{equation}
\frac{\partial\mathbf m}{\partial\phi}
\cdot
\left(
\mathbf m\times\frac{\partial\mathbf m}{\partial z}
\right)
=
\sin\theta\,\theta'.
\end{equation}
Therefore,
\begin{align}
F_{\phi}^{\rm STT}
&=
\frac{P\mu_B j_0}{e}
\int_{-L/2}^{L/2}
\sin\theta\,\theta'\,dz \nonumber\\
&=
\frac{P\mu_B j_0}{e}
\int_{\theta_-}^{\theta_+}
\sin\theta\,d\theta .
\end{align}
Thus
\begin{equation}
F_{\phi}^{\rm STT}
=
\frac{P\mu_B j_0}{e}
\left[
\cos\theta_- - \cos\theta_+
\right].
\end{equation}
Using half-angle identities,
\begin{equation}
\cos\theta_\pm=2c_\pm^2-1,
\end{equation}
one may equivalently write
\begin{equation}
F_{\phi}^{\rm STT}
=
\frac{2P\mu_B j_0}{e}
\left(
c_-^2-c_+^2
\right).
\end{equation}
Finally, the conservative forces are obtained from the energy derivatives. Although they do not enter the equation for the oscillation frequency, they are needed to identify the restoring character of the collective coordinates. For the incomplete wall, the derivative with respect to $\theta_0$, evaluated at the equilibrium widths defined in Eq.~\eqref{eq:si_widths}, is
\begin{equation}
\partial_{\theta_0}U
=
4A
\left[
\frac{s_0}{\delta_1}
-
\frac{s_+^2}{\delta_1s_0}
+
\frac{c_-^2}{\delta_2c_0}
-
\frac{c_0}{\delta_2}
\right].
\end{equation}
For the width coordinate $\delta_1$, the $z>0$ contribution gives
\begin{equation}
\partial_{\delta_1}U
=
\left(
-\frac{4A}{\delta_1^2}
+
2\mu_0H_{\rm Z}M_1
\right)
(c_+ - c_0)
-
\left(
\frac{4A}{\delta_1}
+
2\mu_0H_{\rm Z}M_1\delta_1
\right)
\frac{L}{2\delta_1^2}s_+^2 .
\end{equation}
The expression for $\partial_{\delta_2}U$ is obtained analogously from the $z<0$ branch.

Linearizing the energy derivatives around equilibrium, varying each coordinate separately while holding the others fixed, gives
\begin{equation}
\partial_{\theta_0}U
\simeq
\kappa_{\theta_0}
(\theta_0-\theta_{0,\rm eq}),
\qquad
\partial_{\delta_1}U
\simeq
\kappa_{\delta_1}
(\delta_1-\delta_{1,\rm eq}),
\end{equation}
with positive stiffnesses $\kappa_{\theta_0}$ and $\kappa_{\delta_1}$ in the stable regime. This shows that deviations of $\theta_0$ and $\delta_1$ from their equilibrium values generate restoring forces, justifying their interpretation as confined internal degrees of freedom of the wall.

\section{Maximum current in the linear-profile approximation}

To estimate the maximum current compatible with a pinned configuration, we consider the linear-profile approximation
\begin{equation}
\theta(z)=
\begin{cases}
\pi, & z<0, \\[2mm]
\displaystyle
\pi-\frac{2}{L}(\pi-\theta_+)z,
& 0<z<L/2 .
\end{cases}
\end{equation}
Defining
\begin{equation}
q=\pi-\theta_+\equiv\Delta_+ ,
\end{equation}
the active-side profile becomes
\begin{equation}
\theta(z)=\pi-\frac{2q}{L}z .
\end{equation}

The dissipative coefficient is
\begin{equation}
\Gamma_{\phi\phi}
=
\alpha M_1
\int_0^{L/2}\sin^2\theta\,dz ,
\end{equation}
which gives
\begin{equation}
\Gamma_{\phi\phi}
=
\alpha M_1 L
\left[
\frac14-\frac{\sin(2q)}{8q}
\right].
\end{equation}

The adiabatic STT projection is
\begin{equation}
F_\phi^{\rm STT}
=
\frac{P\mu_B j_0}{e}
\int_0^{L/2}
\sin\theta\,\theta' dz ,
\end{equation}
yielding
\begin{equation}
F_\phi^{\rm STT}
=
-\frac{P\mu_B j_0}{e}
(1-\cos q).
\end{equation}

The gyrotropic coefficient is
\begin{equation}
G_{\theta_+\phi}
=
M_1
\int_0^{L/2}
\sin\theta
\frac{\partial\theta}{\partial\theta_+}
dz ,
\end{equation}
which evaluates to
\begin{equation}
G_{\theta_+\phi}
=
\frac{M_1 L}{2q^2}
\left(
\sin q-q\cos q
\right).
\end{equation}

In the high-field regime considered here, the dynamics is dominated by the Zeeman contribution. We therefore neglect the exchange contribution in the final force balance and retain only the Zeeman energy,
\begin{equation}
U_Z=
-\mu_0H_{\rm Z}M_1
\int_0^{L/2}
\cos\theta\,dz
=
\frac{\mu_0H_{\rm Z}M_1L}{2}
\frac{\sin q}{q}.
\end{equation}
Its derivative is
\begin{equation}
\frac{dU_Z}{d\theta_+}
=
\frac{\mu_0H_{\rm Z}M_1L}{2}
\frac{\sin q-q\cos q}{q^2}.
\end{equation}

The steady-state condition follows from the Thiele equation,
\begin{equation}
G_{\theta_+\phi}
\frac{F_\phi^{\rm STT}}{\Gamma_{\phi\phi}}
=
-\gamma
\frac{dU_Z}{d\theta_+}.
\end{equation}

To determine the maximum current, we consider the limit $q=\pi-\theta_+\rightarrow 0$,
corresponding to the onset of the fully collinear state. Expanding the different terms,
\begin{equation}
G_{\theta_+\phi}
\simeq
\frac{M_1 L}{6}q,
\qquad
\Gamma_{\phi\phi}
\simeq
\frac{\alpha M_1 L}{6}q^2,
\end{equation}
\begin{equation}
F_\phi^{\rm STT}
\simeq
-\frac{P\mu_B j_0}{2e}q^2,
\qquad
\frac{dU_Z}{d\theta_+}
\simeq
\frac{\mu_0H_{\rm Z}M_1L}{6}q .
\end{equation}

Substituting into the steady-state condition, the maximum current density ($j_{\rm max}$) can be found as
\begin{equation}
j_{\rm max}
=
\frac{e}{P\mu_B}
\frac{\alpha\gamma\mu_0H_{\rm Z}M_1L}{3}.
\end{equation}

Finally, the limiting precession frequency within this approximation ($f_{\rm max}$) is obtained from Eq. (5) of the main text and reads
\begin{equation}
f_{\rm max}
=
\frac{\gamma\mu_0H_{\rm Z}}{2\pi}.
\end{equation}

\clearpage
\section{Simulation parameters}
\label{sec:simulation_parameters}

The system is discretized into a one-dimensional chain of $N=50$ spins with spacing $a=0.3\,\mathrm{nm}$, divided into two regions of 25 spins each. The nominal chain thickness is $L=Na=15\,\mathrm{nm}$. The effective magnetization is $M_1=4.0\times10^4\,\mathrm{A/m}$ in the top region and $-M_2=-6.0\times10^4\,\mathrm{A/m}$ in the bottom region. The exchange stiffness, damping, and gyromagnetic ratio are uniform along the chain. Both the perpendicular and cubic anisotropy constants are set to zero. 
\begin{table}[h]
\caption{Common parameters used in the simulations. The constants are listed at the numerical precision used in the calculations.}
\label{tab:simulation_parameters}
\begin{ruledtabular}
\begin{tabular}{lll}
Parameter & Symbol & Value \\
\hline
Number of spins & $N$ & $50$ \\
Spins in each region & $N_1=N_2$ & $25$ \\
Spin spacing & $a$ & $0.3\,\mathrm{nm}$ \\
Nominal chain thickness & $L=Na$ & $15\,\mathrm{nm}$ \\
Exchange stiffness & $A$ & $3.0\times10^{-13}\,\mathrm{J/m}$ \\
Top-region magnetization magnitude & $M_1$ & $4.0\times10^4\,\mathrm{A/m}$ \\
Bottom-region magnetization magnitude & $M_2$ & $6.0\times10^4\,\mathrm{A/m}$ \\
Gilbert damping & $\alpha$ & $0.5$ \\
Gyromagnetic ratio & $\gamma$ & $1.761\times10^{11}\,\mathrm{rad\,s^{-1}\,T^{-1}}$ \\
Spin polarization & $P$ & $0.35$ \\
Perpendicular anisotropy constant & $K_{\rm PMA}$ & $0\,\mathrm{J/m^3}$ \\
Cubic anisotropy & $K_{\rm c}$ & $0\,\mathrm{J/m^3}$ \\
Vacuum permeability (numerical value) & $\mu_0$ & $1.25\times10^{-6}\,\mathrm{H/m}$ \\
Elementary charge (numerical value) & $e$ & $1.60\times10^{-19}\,\mathrm{C}$ \\
Bohr magneton & $\mu_{\rm B}$ & $9.274\times10^{-24}\,\mathrm{J/T}$ \\
\end{tabular}
\end{ruledtabular}
\end{table}

The field and current ranges of the simulations discussed in the manuscript are summarized in Table~\ref{tab:simulation_sweeps}.

\begin{table}[h]
\caption{Field and current ranges considered in the simulations discussed in the manuscript. The number of field or current samples is indicated in parentheses. The static series have STT disabled.}
\label{tab:simulation_sweeps}
\begin{ruledtabular}
\begin{tabular}{llll}
Series &  $\mu_0H_{\rm Z}$ ($\mathrm{T}$) & $j_0$ ($\mathrm{A/m^2}$) \\
\hline
Static field sweep (Fig. 1b)  & $0$--$2$ & 0 \\
Static DW profile (Fig. 1c)  & $8.4$ & 0 \\
Low STT, field sweep (Fig.2c)   & $3$--$30$ & $10^{10}$ \\
Low STT, current sweep (Fig. 2d) & $6$ & $0$--$10^{11}$ (10) \\
Strong STT, current sweep (Fig. 3c-d) & $4$ & $0$--$5\times10^{12}$ (20) \\
\end{tabular}
\end{ruledtabular}
\end{table}

The integration options specify a relative tolerance of $10^{-3}$, an absolute tolerance of $10^{-5}$, and a maximum allowed step of $1\,\mathrm{ns}$. For the low-STT calculations, the current turn-on time is $t_0=10\,\mathrm{ns}$ and the post-turn-on interval extends to $15\,\mathrm{ns}$, giving an output spacing of $5\,\mathrm{ps}$. For the strong-STT current sweep, the corresponding times are $t_0=5\,\mathrm{ns}$ and $8\,\mathrm{ns}$, giving an output spacing of $3\,\mathrm{ps}$. These output spacings are distinct from the internal integration steps.